\documentclass[]{spie}  

\usepackage{amsmath,amsfonts,amssymb}
\usepackage{graphicx}
\usepackage{subcaption}
\usepackage{blindtext}
\usepackage[colorlinks=true, allcolors=blue]{hyperref}
\usepackage{array}
\usepackage{hyperref}
\usepackage{acronym}
\usepackage[squaren, cdot]{SIunits}
\usepackage[skip=6pt]{caption}

\title{MKID, an energy sensitive superconducting detector for the next generation of XAO}

\author[a]{Aurélie Magniez}
\author[a]{Lisa Bardou}
\author[a]{Tim Morris}
\author[a]{Kieran O'Brien}
\affil[a]{Durham University, Department of Physics, South Road, Durham, DH1 3LE, UK}

\newacro{PWFS}{pyramid wavefront sensor}
\newacro{PSF}{point spread function}
\newacro{AO}{adaptive optic}
\newacro{MKID}{microwave kinetic inductance detector}
\newacro{DM}{deformable mirror}
\newacro{MVM}{matrix-vector multiplication}
\newacro{WFS}{wavefront sensor}

\authorinfo{Further author information: (Send correspondence to A. Magniez)\\ E-mail: aurelie.magniez@durham.ac.uk}

\begin{document} 
\maketitle

\begin{abstract}
Selected for the next generation of adaptive optics (AO) systems, the pyramid wavefront sensor (PWFS) is recognised for its closed AO loop performance. As new technologies are emerging, it is necessary to explore new methods to improve it. Microwave Kinetic Inductance Detectors (MKID) are photon-counting devices that measure the arrival time and energy of each incident photon, providing new capabilities over existing detectors and significant AO performance benefits. After developing a multi-wavelength PWFS simulation, we study the benefits of using an energy sensitive detector, analyse the PWFS performance according to wavelength and explore the possibility of using fainter natural guide stars by widening the bandpass of the wavefront sensor.
\end{abstract}

\keywords{Extreme adaptive optics, \ac{MKID}, Pyramid wavefront sensor}

\section{INTRODUCTION}
The \ac{PWFS} \cite{ragazzoni_pupil_1996} is widely recognized as being able to provide the best closed-loop \ac{AO} performance, with many current and future AO systems selecting the \ac{PWFS} as their primary natural guide star wavefront sensor. Existing CCD/CMOS detector technology is well suited to \ac{PWFS} operation, providing near-zero read noise detectors with frame rates of 1 - 3~kHz at either visible or near-infrared wavelengths. However, there is little scope for significant improvement of these detector technologies to enhance \ac{PWFS} AO performance further, or increase sky coverage by observing fainter targets. In the near future, extreme AO (XAO) systems deployed on large telescopes with primary aperture diameters of 20 - 40~m will require high pixel count devices that are capable of high frame rate, low read noise operation. Whilst development work increasing the pixel counts of existing detector technologies is underway \cite{hardy_instrument_2020}, the initial designs for the ELT XAO system\cite{kasper_pcs_2021} used 4 parallel detectors to provide a sufficient number of pixels to control the high-order deformable mirror needed.

Here we propose the use of a microwave kinetic inductance detector (MKID) \cite{day_broadband_2003, mazin_optical_2019} array as an alternative \ac{PWFS} detector technology and describe the benefits this can bring to future AO system performance.

An \ac{MKID} array is a superconducting detector with unique properties compared to CCD or CMOS detectors commonly used as detectors for wavefront sensors. A 2-dimensional array of \ac{MKID} pixels can measure the position, time, and energy of every photon incident on the array. These properties provide a range of new capabilities that an \ac{MKID}-based \ac{PWFS} can take advantage of:
\begin{itemize}
    \item The superconducting properties ensure that the energy of the incident photon is sufficient to avoid any false count thus there is no readout noise.
    \item Each pixel has an independent 1 MHz sample rate readout channel enabling frame rates far in excess of existing \ac{PWFS} detectors and enable new readout modes, for example generating a fixed signal-to-noise frame from the last 10000 photons that updates whenever 100 new photons arrive.
    \item Devices can be made sensitive across visible and near infrared wavelengths simultaneously.
    \item A single device can provide wavelength separated frames to account for wavelength-dependent \ac{PWFS} optical gains or assist with segment co-phasing.
    \item \ac{MKID} fabrication and readout allows freedom in the size, shape, and layout of pixels within the array.
\end{itemize}

In the following sections we first discuss chromatic effects that should be considered when operating a broadband \ac{PWFS}, looking at the optical performance of two pyramids designed for 8~m and 40~m telescope diameters. We then describe the simulation tool that has been developed to investigate some of the unique properties of an \ac{MKID}-based \ac{PWFS}. Next we look at the performance of an AO system using an \ac{MKID}-based \ac{PWFS} sensing two widely separated wavelengths which are outside the standard sensitivity range of existing single detectors. Finally we highlight some of the potential benefits of using an energy sensitive detector within a \ac{PWFS} and highlight future studies that may enhance wavefront sensing performance further.

\section{Optical Dispersion}
\label{sec:optical dispersion}



\begin{figure}
    \centering
    \includegraphics[width=12cm]{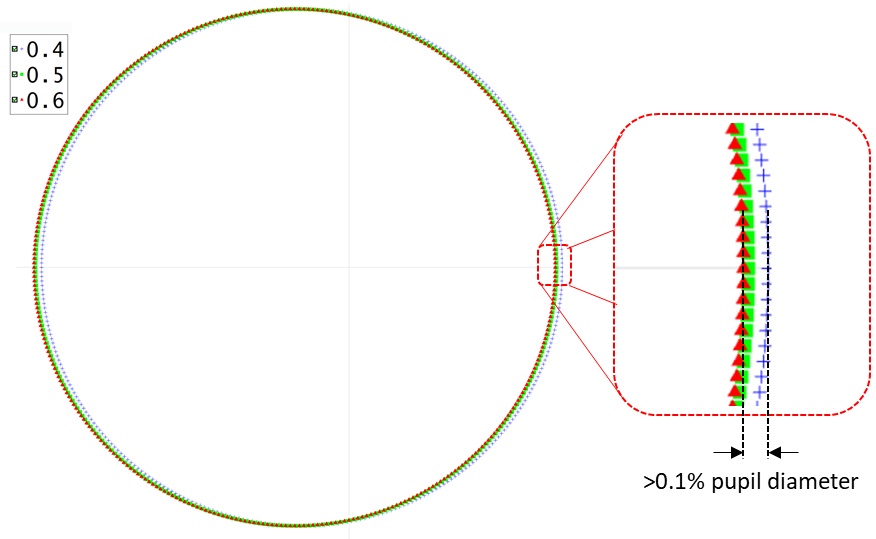}
    \caption{Maximum chromatic pupil offset for the HARMONI\cite{schwartz_design_nodate} \ac{PWFS} operating between wavelengths of 0.4 and 0.6~$\mu m$.}
    \label{fig:dispersion}
\end{figure}

The pyramidal prisms used within the \ac{PWFS} introduce unavoidable chromatic dispersion at the pupil. A double pyramid configuration using multiple glasses can reduce the amount of chromatic pupil dispersion \cite{tozzi_double_2008}, but it is difficult to reduce chromatic pupil dispersion to zero over a wide waveband. In this section, we characterise the pupil dispersion across the wavelength range from 0.4 to 2.2 microns for the HARMONI \ac{PWFS} design \cite{schwartz_design_nodate} to determine the spectral bandpass that can be achieved with a \ac{PWFS} designed for ELT use before chromatic effects may need to be be considered. There have no detailed studies investigating the impact of wide spectral bandpass \ac{PWFS} performance, however Schwartz \textit{et al.}\cite{schwartz_design_nodate} shows that the HARMONI \ac{PWFS} performance degrades when the pupil is offset by 0.2 \ac{PWFS} pixels. Whilst the AO system performance may not be as sensitive to the blur introduced by a chromatic pupil dispersion as opposed to the pupil offset studied by Schwartz \textit{et al.}, a full study of the impact of chromatic pupil dispersion on system performance is outside the scope of this proceeding. Here we simply adopt a $\pm 0.1$~pixel offset as a suitable limit where we could expect \ac{PWFS} performance to start to degrade \cite{heritier_new_2018}. Figure~\ref{fig:dispersion} shows the dispersion of the HARMONI \ac{PWFS} optics between the wavelengths of 0.4 and 0.6~\micro\metre. Note that this is outside the design and operating range of the HARMONI \ac{PWFS}, but serves as an example here only.

The following results were generated using a Zemax model (version 29/10/2021) of the HARMONI and LBT \ac{PWFS}. The HARMONI pyramid fed by a telecentric paraxial $f/30$ beam, and the LBT Pyramid by a non-telecentric f/35 beam. We have assumed a 30 pixel per pupil sampling for the LBT pyramid \ac{PWFS} and a sampling of 250 pixels across the HARMONI pyramid. This is higher than the 100 pixel sampling used in HARMONI and broadly equivalent to the proposed sampling needed for XAO \ac{PWFS} for PCS \cite{kasper_roadmap_2013}. 

\begin{figure}
    \centering
    \includegraphics[width=\textwidth]{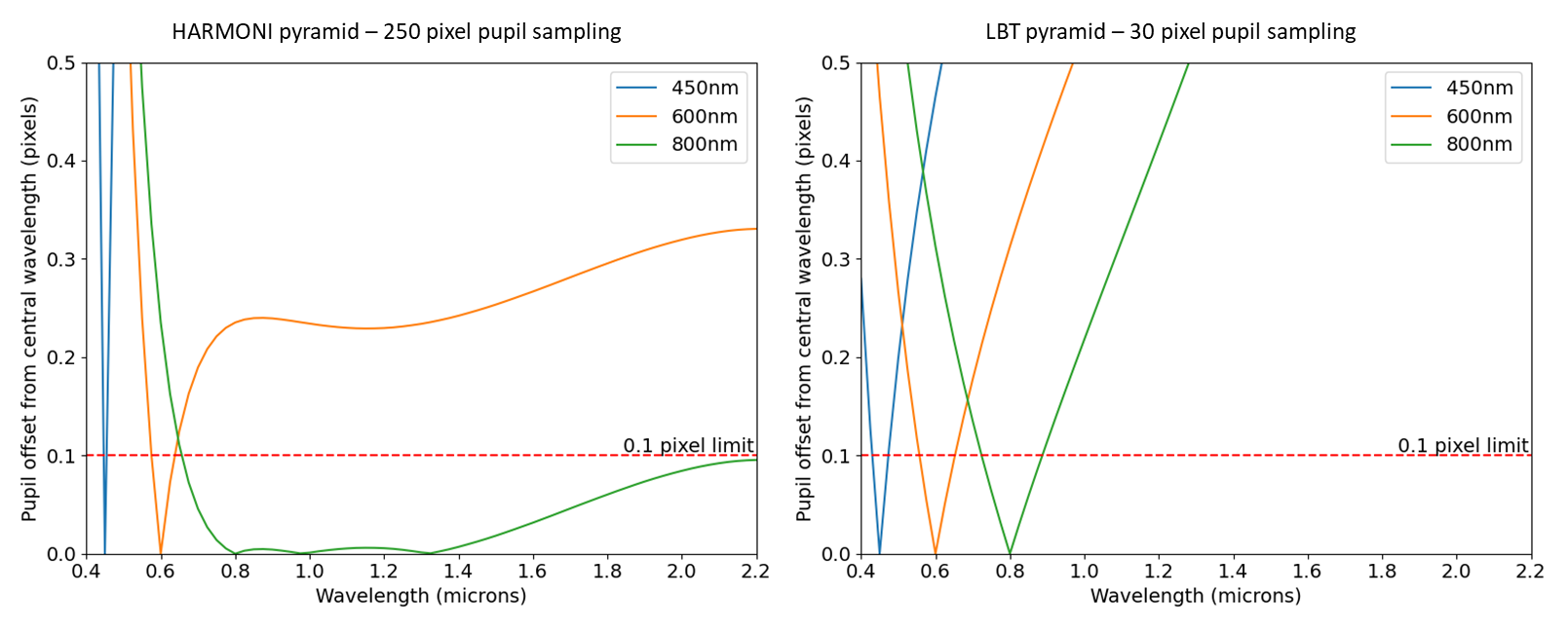}
    \caption{Maximum chromatic pupil dispersion across the wavelength range of 0.4 to 2.2~$\mu m$ for the HARMONI \ac{PWFS} sampled by 205 pixels per pupil (left) and LBT \ac{PWFS} sampled by 30 pixels per pupil (right) operating at three central wavelengths of 450, 600 and 800~nm.}
    \label{fig:dispersion_plot}
\end{figure}

From Figure \ref{fig:dispersion_plot} we can see that the HARMONI pyramid design does achieve a level of residual pupil dispersion well below the 0.1 pixel across a wavelength range of 0.7 to 2.2~$ \mu m$, even for the 250 pixel pupil ELT-scale XAO sampling considered here. However when operating at central wavelengths of less than 0.8~$\mu m$, the bandpass of the HARMONI \ac{PWFS} design where chromatic pupil dispersion can be ignored can be restricted to a few tens of nanometers bandpass only. Figure \ref{fig:dispersion_plot} shows that the chromatic pupil dispersion of the LBT \ac{PWFS} is much higher, restricting operation to approximately $\pm 0.1$ of the the central wavelength of \ac{PWFS} operation for the 30 pixel pupil sampling. It is not surprising that a pyramid designed for an 8~m class telescope does not match the performance of a pyramid designed for a 40~m telescope and there are undoubtedly modifications that could be made to either design to optimise over a wider spectral bandpass whilst keeping dispersion at acceptable levels. However these plots do indicate that a \ac{PWFS} detector capable of spectrally resolving wavelength at the level of R=10-30 at visible wavelength could be beneficial, particularly for XAO-scale wavefront sensing across wide spectral bandpass. We note that upcoming high pixel count MKID arrays are expected to achieve spectral resolution of R=30\cite{zobrist_membraneless_2022}
, and current low pixel count arrays are achieving a spectral resolution of R=50\cite{de_visser_phonon-trapping-enhanced_2021}, beyond the required spectral resolution suggested here. Spectral resolution and pyramid chromatic dispersion are undoubtedly \ac{WFS} design trade offs that could be made to optimise system performance over the wide passbands considered in this work. We also note that some amount of chromatic pupil dispersion coupled to the wavelength sensitivity may enable techniques such as super-resolution \cite{fusco_key_2022}.



\section{Simulation}
\label{sec:Simulation}

SoaPy \cite{marchetti_soapy_2016}, available on Github : \url{https://github.com/AOtools/soapy}, is an end-to-end adaptive optics python simulation that handles atmosphere, telescope, deformable mirrors and wavefront reconstruction. Throughout this section, we describe the new \ac{PWFS} module that has been added to Soapy. In addition to the standard simulation appropriate for a CCD-based \ac{PWFS} model, there are two features specific to an \ac{MKID}-based \ac{PWFS} that can be implemented in the simulation. First, wavelength scaling is needed to analyse \ac{PWFS} sensing at multiple wavelengths simultaneously and second, the photon stream must be recreated, recreating the ability of \ac{MKID} to detect the arrival time of each individual photon. Recreating the photon stream requires running the simulation with smaller time steps than the modulation frequency, with the ability to vary the modulation path at the Pyramid tip on a frame-by-frame basis. However, in this work the focus is on the multi-wavelength aspect and there will be no further discussion of the photon stream.

\subsection{PWFS module description}

\begin{figure}
    \centering
    \includegraphics[width=\textwidth]{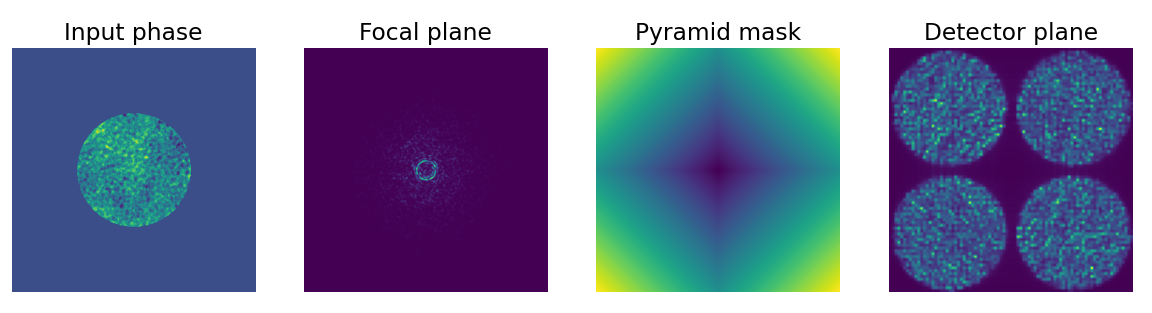}
    \caption{Illustration of the different steps of the PWFS module using the parameter shown in Table \ref{tab:Sim_parameters} scaled at 500 nm for a modulation radius of 0.08 arcseconds. From left to right : Input phase of the PWFS after the loop had been closed for 10 frames, Modulated PSF, Noiseless detector frames }
    \label{fig:PWFS module}
\end{figure}

To implement the \ac{PWFS} module, we follow the standard numerical model approach described in Verinaud \emph{et al.}\cite{verinaud_nature_2004} as a basis. We first generate a modulated \ac{PSF} at the tip of a four-sided pyramidal prism, with modulation of a defined path, in this case circular, introduced by a tip-tilt mirror. The pyramidal prism divides the beam into four, and a re-imaging lens is used to project the resulting four pupil images onto the detector. Implementing the energy sensitivity of an MKID within an end-to-end simulation requires additional functionality and imposes constraints on the simulation parameters. We reproduce energy sensitivity by simulating several monochromatic \ac{PWFS} across the wavelength range of interest, whilst ensuring that the modulation amplitude and pupil sampling is consistent between each \ac{PWFS}.   Each step is represented by the following model illustrated Figure \ref{fig:PWFS module}, \ac{PWFS}\(_{\lambda}\) is the scaled \ac{PWFS} model for a wavelength \(\lambda\) :

\begin{enumerate}
    \item A random atmospheric 2-dimensional phase screen is generated with a defined r$_0$ at 500~nm, and the phase sampled by the telescope aperture at a given time step is sliced from this screen. The phase screen sampling is set to ensure that the corresponding field of view in the focal plane at the tip of the pyramid prism is wide enough not to truncate the signal. It is also set so that the number of pixels across the pupil is a multiple of the number of pixels across the pupil images on the detector. This latter constraint is included to avoid interpolation errors when generating the final \ac{PWFS} image. The array containing the phase screen is then sufficiently padded to make sure that the \ac{PSF} in the focal plane is Nyquist sampled. The atmosphere aberrations are expressed in path difference first and then translated in phase corresponding to each wavelength.
    \item A tip-tilt phase screen is added to the atmospheric phase  to model the modulation. The modulation amplitude, as a parameter of the simulation, is defined in arcseconds (on-sky) and not in $\lambda$/D units because the modulation radius would then be different for each PWFS$_{\lambda}$. This means that amplitude of the tip-tilt phase screen has to be scaled by the central wavelength of each \ac{PWFS}$_{\lambda}$.   
    \item We perform the Fourier transform of the complex amplitude defined by the modulated atmospheric phase and the telescope aperture to obtain the illumination pattern in the focal plane at the top of the pyramid.
    \item The complex amplitude of a phase mask recreating the phase profile of the pyramid is generated and multiplied by the complex amplitude of the modulated \ac{PSF}. The amplitude of each tilted phase physically represents the prism tip angle and is chosen to give the desired pupil separation at the detector plane.
    \item The square of the absolute value of the inverse Fourier transform is computed to obtain the intensity of the pupil images on the detector. It is important to note that the only parameters determining the pupil images position and shape are the number of phase pixels across the pupil and the pupil separation introduced by the pyramid phase mask . 

\end{enumerate}

The extracted slopes or the x and y axes respectively are then obtained using the following formula \cite{ragazzoni_pupil_1996} :

\begin{equation}
    S_x(x,y) = \frac{(I_1(x,y) + I_2(x,y))-(I_3(x,y) + I_4(x,y)) }{\Sigma I}  
    \label{eq:Sx}
\end{equation}

\begin{equation}
    S_y(x,y) = \frac{(I_1(x,y) + I_3(x,y))-(I_2(x,y) + I_4(x,y)) }{\Sigma I}
    \label{eq:Sy}
\end{equation}

With \(I_i(x,y)\) the intensity of the pixel (x,y) of the i-image of the pupil on the detector. 

The multiple \ac{PWFS} images can then either be processed independently to emulate the energy sensitivity of the \ac{MKID}, or summed to emulate a CCD operating over this same passband with an achromatic optical system. The SoaPy \ac{PWFS} module does not yet contain the functionality to simulate the exact spectral resolution of the \ac{MKID}, which introduces some uncertainty on the energy of any given detected photon but this is planned for a future upgrade. 

\subsection{Parameters of the simulation}
\label{sec:Parameters}

Prior to analysis, we must first determine the parameters of the baseline simulation, presented in Table \ref{tab:Sim_parameters}. To investigate performance in a realistic configuration we based our telescope diameter, frame rate, imaging wavelength and pupil sampling to match that of SPHERE\cite{fusco_high-order_2006}. In order to simplify the simulation we use a DM that replicates Karhunen-Loeve modes\cite{cannon_optimal_1996} rather than a realistic model including actuators and their influence functions. We use the DM to control 500 modes. The number of modes and closed loop gain was chosen to achieve reasonable \ac{AO} correction  while keeping the number of modes small enough to avoid loop instabilities. We set our loop with a one frame delay, which is unrealistic but has the advantage of not complicating the analysis with an additional temporal error. The pupil separation is fixed to be enough to properly separate the pupil images while remaining as small as possible in order to reduce simulation array sizes. As an MKID has no readout noise, it is set to 0. The rest of the simulation parameters are chosen according to the wavelength studied and discussed in the following section.

\begin{table}[h]
    \centering
    \setlength{\tabcolsep}{10pt} 
    \renewcommand{\arraystretch}{1.5} 
    \begin{tabular}{|>{\centering\arraybackslash}m{0.35\textwidth} |
    >{\centering\arraybackslash}m{0.1\textwidth} | }
        \hline
         Parameter & Value  \\[0.5ex] 
         \hline
         \hline
         Telescope Diameter & 8 m \\
         Frame rate & 1 kHz\\
         Imaging wavelength & 1.65 $\mu$m\\
         PWFS pupil sampling & 40 pixels \\
         Pupil separation & 4 pixels\\
         \ac{DM} Actuators & 500\\
         Closed-loop gain & 0.6\\
         Latency & 1 frame\\
         Read noise & 0\\
         \hline
    \end{tabular}
    \caption{List of the parameters used by the \ac{PWFS} model throughout this proceeding}
    \label{tab:Sim_parameters}
\end{table}

\subsection{PWFS behaviour}
\label{sec:linearity}
The PWFS shows a non linear behaviour that can be mitigated by adding a modulation of the \ac{PSF} on tip of the pyramidal prism. Increasing this modulation leads to an increase of the dynamic range at the price of a loss of sensitivity of the measurement. This is the well-known trade-off between sensitivity and dynamic range\cite{ragazzoni_pyramid_2002,verinaud_nature_2004}. One of the strength of the \ac{PWFS} is we can tune the modulation according to the strength of the atmospheric perturbations once on sky.

Another feature of the \ac{PWFS} is a change of behaviour depending on the characteristics of the incoming perturbation. In practice, as long as the \ac{PWFS} operates within its linear range, it translates into a loss in the measurement of a given signal between calibration and on-sky operation. Optical gains\cite{korkiakoski_improving_2008} are the metric used to quantify this loss. This effect is particularly problematic, as optical gains change with the r$_0$ which itself varies when operating on sky. It exist different methods to compensate them\cite{deo_telescope-ready_2019, chambouleyron_pyramid_2020-1} and implementing them in our simulation is planned in a future work. 

\section{Two wavelengths example}

Whilst sensing at two wavelength is achievable using dichroic and two independent cameras, extending this beyond two wavelengths rapidly becomes optically and mechanically complex and costly. With an \ac{MKID}, a \ac{PWFS} could be used to differentiate multiple wavebands across a wide passband using  the same detector. One could then imagine a complete division of the detected spectrum into several colours. The question is then what interesting properties could this chromatic distinction bring? In order to better understand the response of the \ac{PWFS} as a function of wavelength, we will focus on studying the behaviour of two wavelengths: 500~nm and 1700~nm.

The choice of these wavelengths covers the expected wavelength range possible with an \ac{MKID}-based \ac{PWFS}. At shorter wavelengths, light would start to be absorbed by the atmosphere and at longer wavelengths the sky background would begin to dominate without cooling the rest of the optical system.  It is also worth noting that their observation would not be optimal or even possible by a single CCD across this passband, and the wide wavelength sensitivity of the \ac{MKID} allows us to take advantage of the possibility of observing in the infrared and the visible simultaneously.

The modulation amplitude \cite{verinaud_nature_2004} has an impact that is relative to the wavelength. It is therefore necessary to choose it so that both PWFS$_{500}$ and PWFS$_{1700}$ can achieve good performance. Thus the amplitude modulation is selected according to the trade-off between sensitivity using small modulation, and linearity using large modulation. Bond \emph{et al.} have shown that modulation radii of $2$ to $4 \lambda/D$ allow to reach good performance for a \ac{PWFS} operating in the H-band. Choosing those values for 1700~nm corresponds to modulation radii of $6$ and $12 \lambda/D$ at 500~nm which are large but within reason, therefore we use those values in our simulations. The wavelength-dependent value of the  Fried parameter \cite{fried_statistics_1965} at 500~nm and 1700~nm is shown in Table \ref{tab:Sim_parameters500&1700} along with the modulation radii in $\lambda$/D and on-sky angular units.

\begin{table}[h]
    \centering
    \setlength{\tabcolsep}{10pt} 
    \renewcommand{\arraystretch}{1.5} 
    \begin{tabular}{|>{\centering\arraybackslash}m{0.35\textwidth} |
    >{\centering\arraybackslash}m{0.1\textwidth} | 
    >{\centering\arraybackslash}m{0.1\textwidth} | 
    >{\centering\arraybackslash}m{0.1\textwidth} | }
        \hline
         Parameter & Fixed values & 500 nm & 1700 nm  \\[0.5ex] 
         \hline
         \hline
         Fried Parameter $r_0$ & & 0.12 m & 0.33 m \\
         Modulation radius & 0.08 arcsec & 6$\lambda$/D  & 2$\lambda$/D \\
          & 0.16 arcsec & 12$\lambda$/D  & 4$\lambda$/D \\

         \hline
    \end{tabular}
    \caption{ List of the parameters used by \ac{PWFS}\(_{500}\) and \ac{PWFS}\(_{1700}\) and their relative values. 0.08 arcseconds is the default modulation amplitude and 0.16 arcseconds is not used in section \ref{sec:MKID rec}}
    \label{tab:Sim_parameters500&1700}
\end{table}

\begin{figure}
    \centering
    \includegraphics[width=\textwidth]{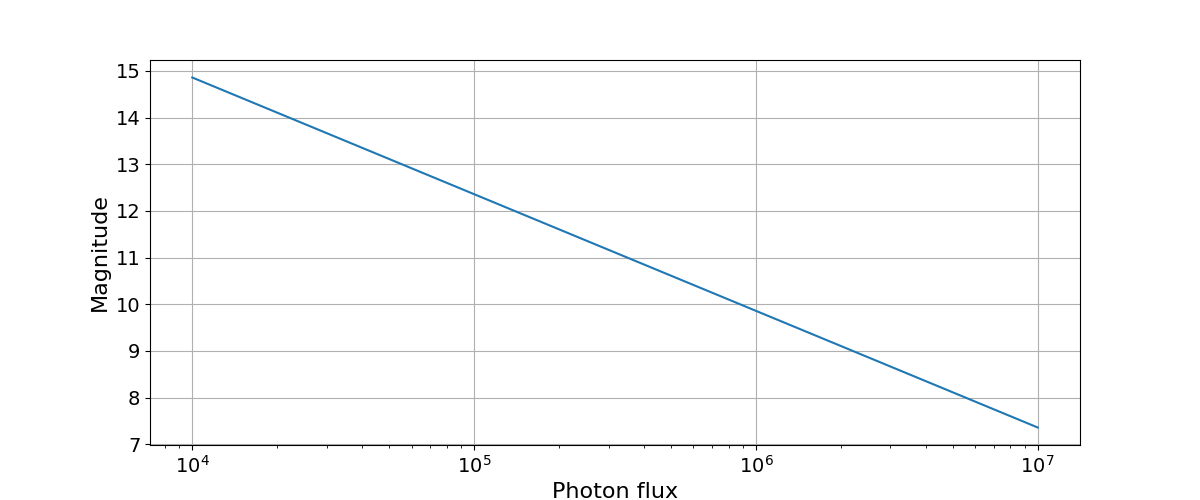}
    \caption{Conversion between magnitude and detected photon flux at the \ac{PWFS}}
    \label{fig:magvsflux}
\end{figure}

One of the goals to use an \ac{MKID} is to have the possibility to use fainter natural guide stars. We are therefore interested in the performance of the \ac{PWFS} as a function of the incoming photon flux or magnitude. For the calculation of the magnitude, we do not consider any particular stellar spectrum but set the photon flux on the PWFS$_{500}$ and PWFS$_{1700}$ to be equal. However, we have seen in the Section \ref{sec:optical dispersion} that the passband around 500~nm where dispersion remains below $\pm 0.1$~pixels can be considered is narrower than the one around 1700~nm. We accept this method of calculating magnitude is a simplification, knowing that for a stellar spectrum the photon flux is typically lower in the infrared than in the visible for most bright guide stars. Figure \ref{fig:magvsflux} shows the conversion between the magnitude and the photon flux used for this section for reference.

\subsection{Comparison of \ac{PWFS}\(_{500}\) and \ac{PWFS}\(_{1700}\) performance}
\label{sec:two wave}

\begin{figure}
    \centering
    \includegraphics[width=\textwidth]{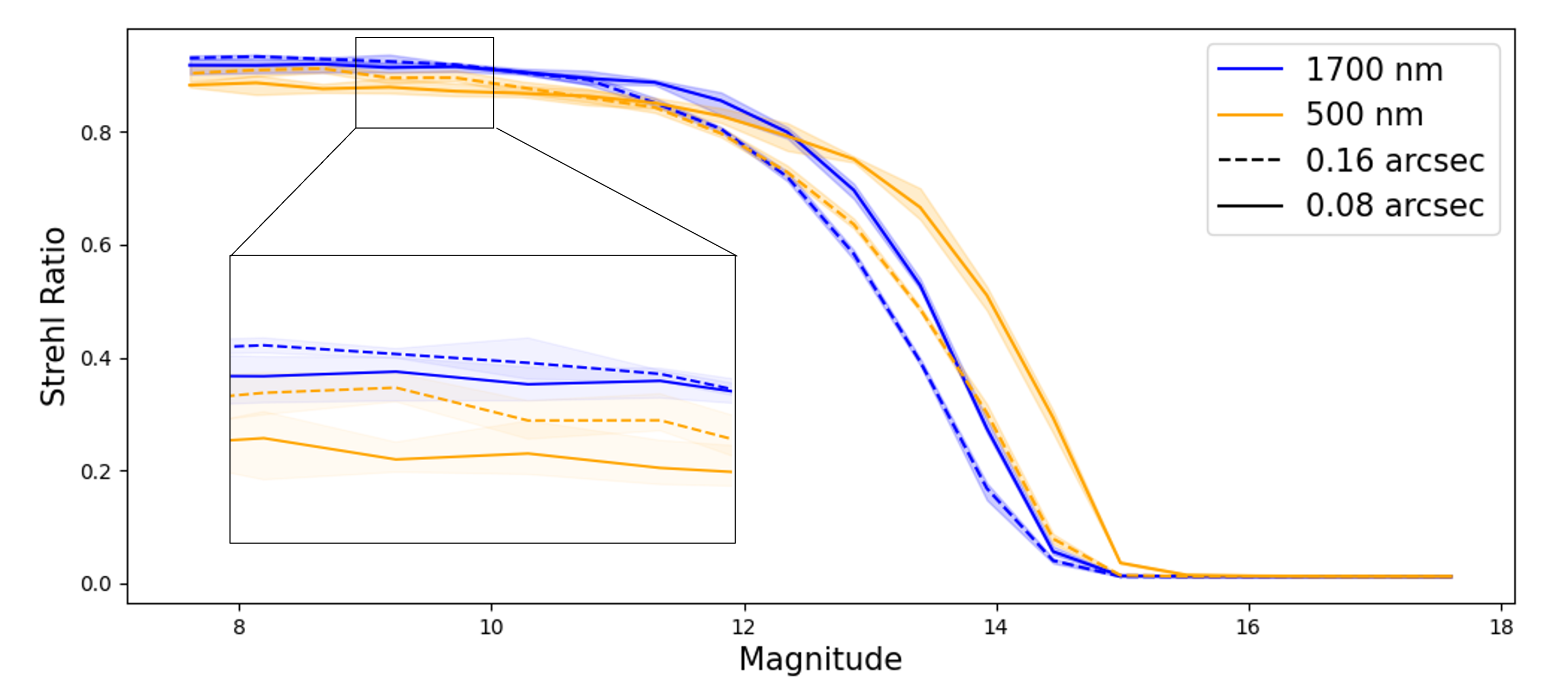}
    \caption{Strehl ratio of a \ac{PWFS} with parameters from Tables \ref{tab:Sim_parameters} and \ref{tab:Sim_parameters500&1700} at 500~nm and 1700~nm for modulation radius of 0.16 and 0.08 arcseconds as a function of guide star magnitude. The curves represent the average of 4 different simulations with 4 atmospheres of the same generation parameters.}
    \label{fig:Strehl at different wvl and modulation}
\end{figure}

To understand how to take advantage of different wavebands, it is first necessary to examine the performance of each \ac{WFS}. In Figure \ref{fig:Strehl at different wvl and modulation}, we look at the Strehl ratio of PWFS$_{500}$ and PWFS$_{1700}$ as a function of magnitude. There are two distinct performance zones. There is a high plateau of the Strehl ratio at small magnitudes, i.e. high photon flux, where the photon noise is negligible. A reduction in performance is  observed for stellar magnitudes above 11 as \ac{WFS} noise starts to dominate.

We observe that at the low magnitude plateau, increasing the modulation amplitude and therefore the linear range improves the performance of the AO system, whereas in the low-photon regime it is more advantageous to have a smaller radius of modulation. We also observe that the \ac{PWFS} performs better at 1700~nm at high flux, but using 500~nm as a sensing wavelength allows to use the \ac{PWFS} at fainter magnitudes. 

\begin{figure}
    \centering
    \includegraphics[width=\textwidth]{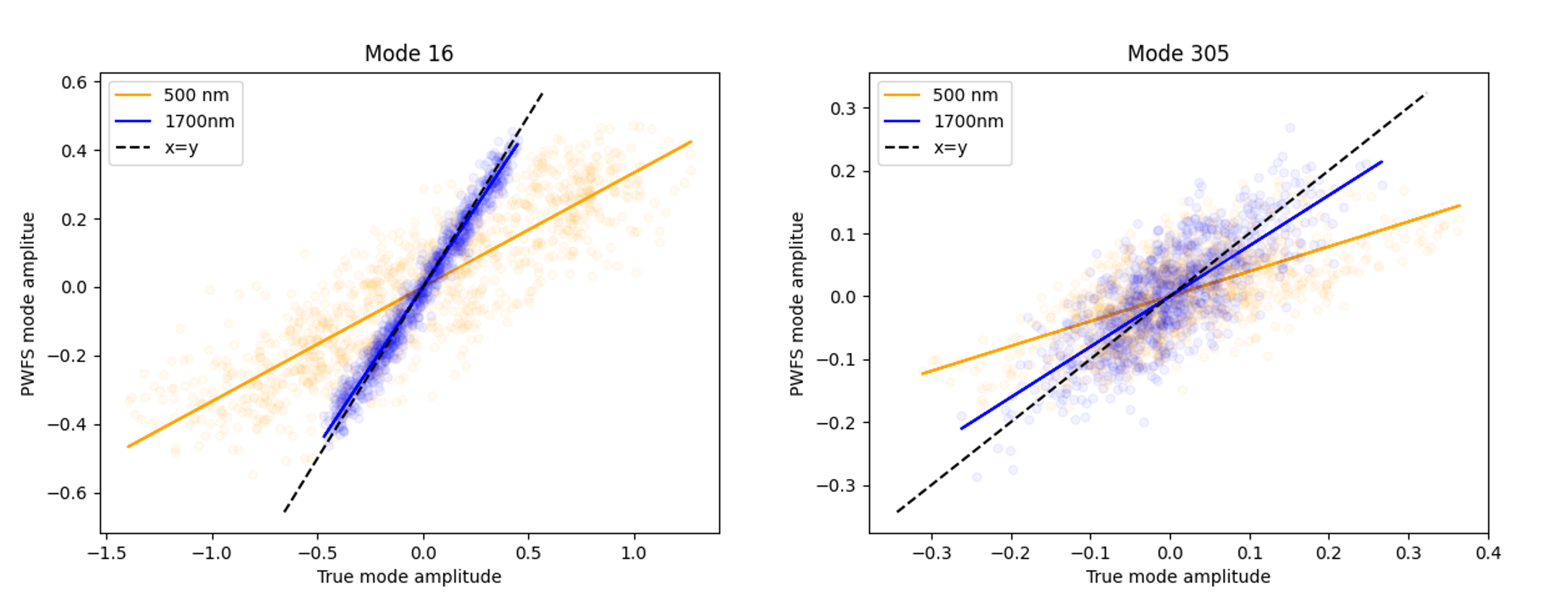}
    \caption{Plot of the input (true) versus \ac{PWFS} measured mode amplitude of the \ac{PWFS} for 2 KL modes at 500 nm and 1700 nm for magnitude 10 guide star.}
    \label{fig:Plot True vs detected}
\end{figure}

To better understand this, we examine the behaviour of optical gains, which correspond to a loss of sensitivity that is a function of both the spatial frequency and residual wavefront aberration\cite{korkiakoski_improving_2008,deo_telescope-ready_2019}. In simulation, these optical gains can be directly obtained by plotting for each mode the input amplitude versus the amplitude detected by the PWFS. An optical gain of 1 would be equivalent to the \ac{PWFS} having the same response on-sky to a given signal as during calibration of the interaction matrix. This is illustrated in Figure \ref{fig:Plot True vs detected}. It can be seen that for a bright magnitude the optical gains are higher at 1700~nm than at 500~nm as expected\cite{bond_adaptive_2020}, which explains that the \ac{PWFS} performs better at 1700~nm  with a high photon flux.

As the magnitudes increases, photon noise begins to dominate and a change in performance is observed. The PWFS$_{500}$ then becomes more efficient. It can be explained by the fact that the amplitudes of atmospheric aberrations expressed as phase are smaller at 1700~nm than at 500~nm, consequently, the signal measured by the \ac{PWFS} - that is the first derivative of the incoming phase - also has smaller amplitude at 1700~nm and therefore is more easily hidden in the photon noise, which is constant between the two sensed bands in this simulation. The difference in amplitude of the incoming signal for each of the wavelengths is particularly visible in the left-hand side plot of Figure~\ref{fig:Plot True vs detected}: the dispersion of value along the x-axis for the simulation at 1700~nm is significantly smaller than at 500~nm. 

In that low flux regime it is also more advantageous to use a smaller radius of modulation. We hypothesise that this is caused by the loss of sensitivity linked to extending the linear range: if the radius of modulation is bigger, a given signal translates into smaller amplitudes on the detector plane which are in turn more easily lost in noise. Further study is necessary to confirm this hypothesis.

Without trying to use both PWFS$_{500}$ and PWFS$_{1700}$ together, it can be seen that if a system was designed with two \ac{PWFS}, one operating at 500~nm and a second at 1700~nm, we would gain approximately a half magnitude gain if we used the short wavelength \ac{PWFS} when observing fainter guide stars.

Figure~\ref{fig:Plot True vs detected} also shows the change in optical gain between wavelength for different modes. We compute a linear fit of the data sets to get their values which are 0.33 at 500~nm and 0.93 at 1700~nm for mode 16 and 0.39 at 500~nm and 0.80 at 1700~nm for mode 305. Whilst the linear fits to each mode/wavelength pairing plotted here exhibit different gradients, we know that these are measuring the same input aberration. The deviation between these gradients will vary dependent on the residual AO correction and modulation amplitude and could provide a real-time estimation of optical gain, but this study is beyond the scope of this paper. 

\subsection{First reconstruction with an \ac{MKID}}
\label{sec:MKID rec}

As we have seen in the Section \ref{sec:optical dispersion}, a dispersion of more than 0.2 pixels leads to a decrease in PWFS performance. With a CCD-based PWFS, we would need to dichroically split the signals onto different cameras to be able to use all the signals efficiently. This optical system, theoretically feasible for a limited number of wavebands, would be very complex to implement as opposed to the use of a single MKID array. The \ac{MKID} would allow us to combine wavelengths to improve performance and observe fainter stars by considering a wider range of wavelengths.

\begin{figure}
    \includegraphics[width=\textwidth]{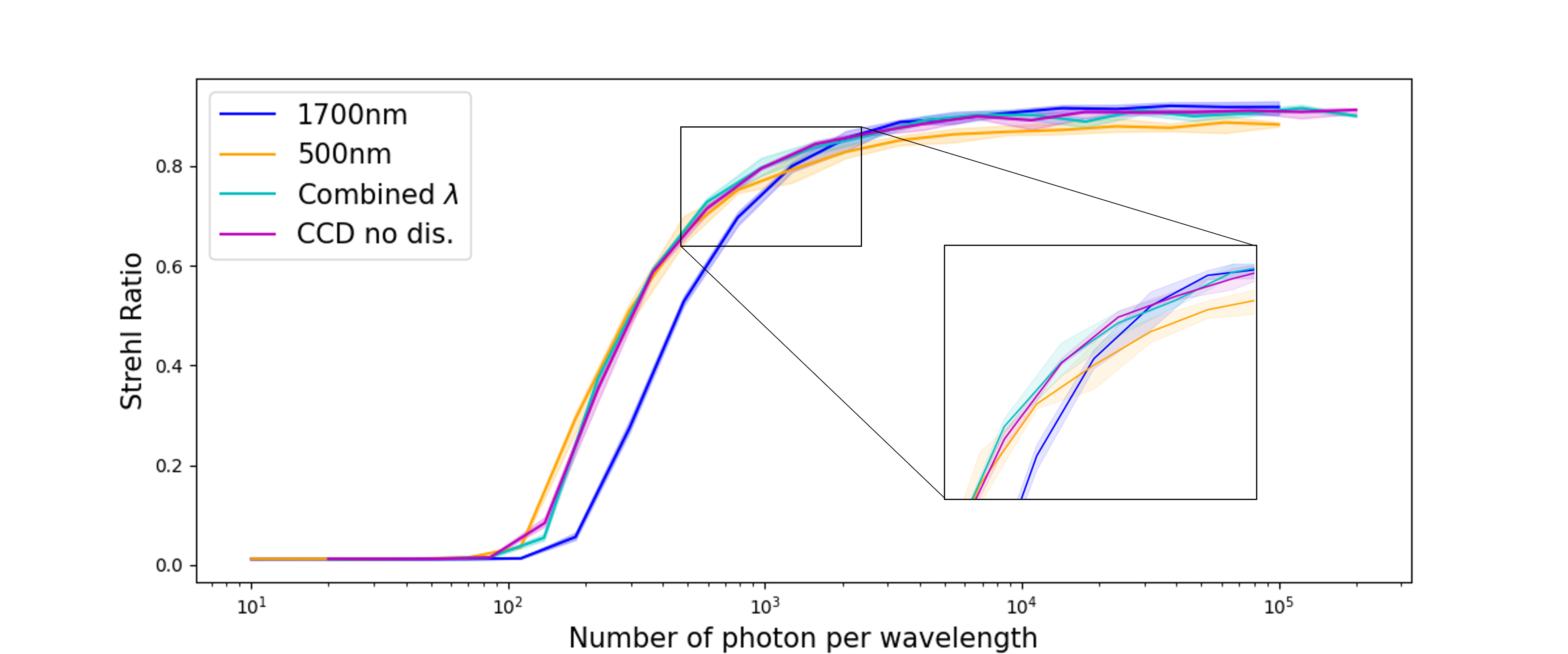}
    \caption{AO system Strehl ratio for the system with parameters shown in Table \ref{tab:Sim_parameters} and a \ac{PWFS} operating at 500 nm and 1700 nm as a function of photon flux.The curves represent the average of 4 different simulations with 4 atmospheres of the same generation parameter}
    \label{fig:phtnumber}
\end{figure}

\begin{figure}
    \centering
    \includegraphics[width=\textwidth]{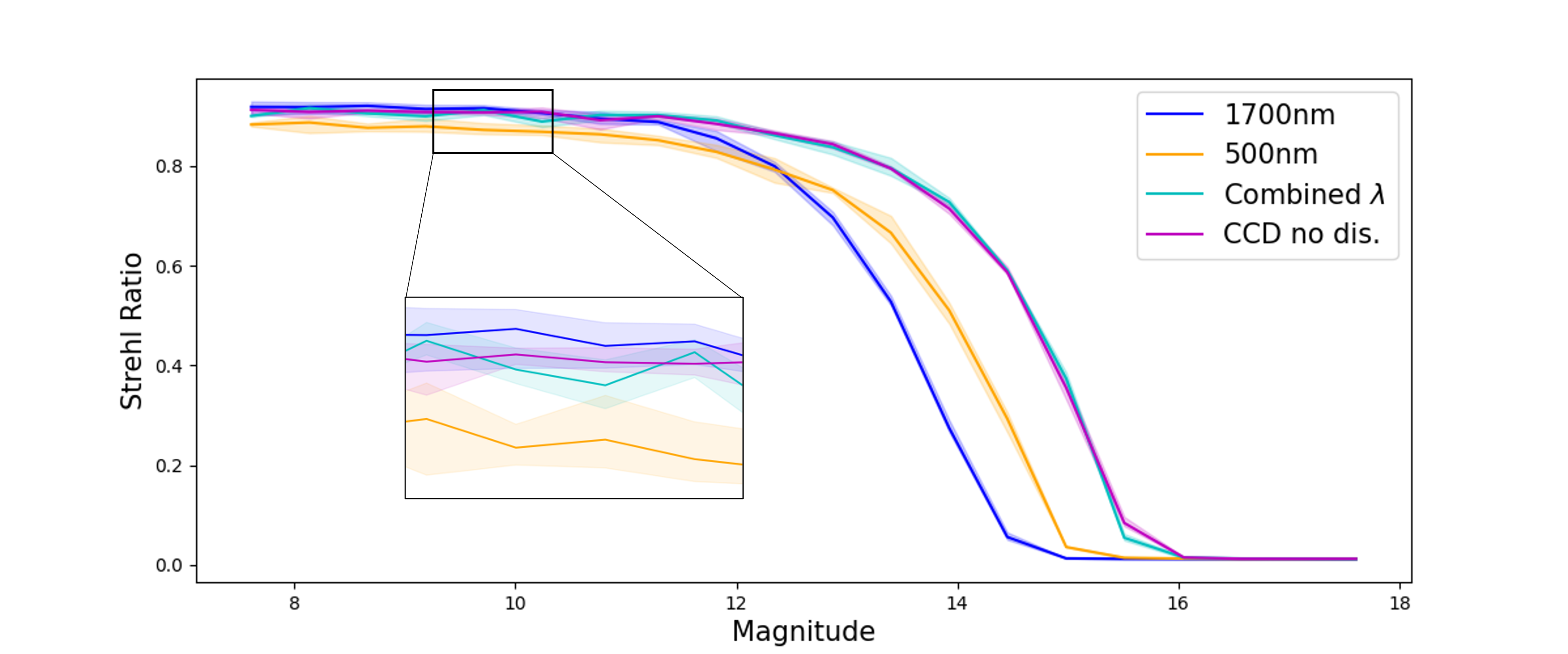}
    \caption{AO system Strehl ratio for the system with parameters shown in Table \ref{tab:Sim_parameters} and a \ac{PWFS} operating at 500 nm and 1700 nm as a function of guide star magnitude. The curves represent the average of 4 different simulations with 4 atmospheres of the same generation parameter}
    \label{fig:Magnitude}
\end{figure}

In Figure \ref{fig:phtnumber} and \ref{fig:Magnitude}, we look at the performance of two simple reconstruction schemes that can be used by an \ac{MKID}-based \ac{PWFS}:\smallskip
\begin{itemize}
\item \textbf{CCD with no dispersion} This algorithm simulates the use of an unrealistic CCD-based \ac{PWFS} which has a wide waveband of detection and a perfect achromatic prism, i.e. no pupil dispersion. To do this, we add the intensities of PWFS$_{500}$ and PWFS$_{1700}$. The \ac{PWFS} module in the simulation generates pupil images that are perfectly overlapping with the same pupil sampling. The slopes are then calculated from equations \ref{eq:Sx} and \ref{eq:Sy}. The system control matrix is generated from the pseudo-inverse of the interaction matrix, using singular value decomposition. Modes corresponding to the 42 smallest singular values were zeroed.
\item \textbf{Combined wavelength} This algorithm is specific to an energy sensitive detector. Slopes S$_x$ and S$_y$ of the two PWFS$_{\lambda}$ are calculated independently. They are then concatenated to double the slope size compared to a monochromatic PWFS and the same pseudo-inverse of the interaction matrix is used to generate the control matrix.  This simplistic method deserves further investigation as regards the optimal method for polychromatic reconstruction.
\end{itemize}

While the two simulations have similar performance, it is important to note that the ``CCD with no dispersion" simulation is unrealistic as no single CCD has either a sensitive waveband ranging from 500 to 1700 nm, nor zero readout noise. This line is included as a comparison only to show the operation of the \ac{PWFS} with double the photon flux.

It was seen in Section \ref{sec:two wave} that signals at different wavelengths can be used independently to improve performance. In Figure \ref{fig:phtnumber}, we compare system performance with the two schemes described above for the same number of photons per frame. In both cases, the performance of PWFS$_{\lambda}$ in the transition zone between the high-flux and low-flux regimes is at the level of the best of PWFS$_{500}$ and PWFS$_{1700}$ irrespective of flux. Looking at Figure \ref{fig:Magnitude}, we can see that by combining the two signals, we achieve a gain of one magnitude. By considering the 500nm and 1700nm \ac{PWFS} signals as two independent wavefront sensors it is clear that an \ac{MKID} array could provide a clear improvement of the performance of a system operating only at a single wavelength. This ``combined wavelength" reconstruction is a very first approach and there are several ways of combining signals for each wavelength that we wish to consider in future work that may improve performance further, particularly as we increase the number of sensed bands beyond the two presented in this work.

\section {Conclusion}
\label{sec:conclusion}  
Optimising PWFS performance is a crucial task to maximise the astronomical science performance of the next generation of telescopes and instruments. However, existing wavefront sensors are limited by their detectors and the optical systems that limit their operation to a narrow passband of the stellar spectrum. We have discussed here the possibility to upgrade the PWFS by using an energy-sensitive superconducting detector such as an MKID.

The optical dispersion of a pyramidal prism defines the bandpass of the sensor as a pupil dispersion greater than $\pm~0.1$ pixels degrades the performance of the AO reconstruction. Unless the optical design of a \ac{PWFS} had been fully achromatic ensure a minimum of dispersion over a large waveband such as the HARMONI design, it is possible to go beyond and use a wider bandpass by using an energy-sensitive detector. Existing MKID arrays can provide sufficient energy resolution (up to an R=30) to allow existing \ac{PWFS} designs to extend their operating passband beyond that of their initial design range, thereby increasing detected photon flux.

To analyse the use of the energy sensitivity of the \ac{MKID}-based PWFS, we have implemented a new PWFS module within the Soapy simulation package. The analysis was broken down into a simple case of two wavelengths to understand how we could use wavelength dependent signals. Those wavelengths, 500nm  and 1700 nm, had been chosen to illustrate the edges of the bandpass we would consider with the use of an MKID on existing optical/near infrared telescopes. We saw that at different regimes of photon flux longer wavelengths provide better correction in the high flux regime and shorter wavelengths allow the system to push further towards the faint magnitude limit. The use of an \ac{MKID} makes it possible to differentiate these signals and further optimise performance across a wider \ac{PWFS} operating range. We have shown that by considering the slopes measured at each wavelength independently within the reconstruction, we can achieve better performance in both low and high-flux conditions, also improving performance within the transition zone between these two flux regimes and gaining from the increased flux enabled by the larger bandpass.

We believe that this simple two wavelengths simulation begins to show the benefits in the use of an energy sensitive detector to pyramid wavefront sensing, but we are clear there is large amount of work to fully identify and characterise the potential performance benefits. Future work on this topic will be to investigate the combination of signals from more than the two wavelengths studied here, characterisation of the variation of optical gain with wavelength, and potentially the introduction of chromatic pupil dispersion between wavelengths to allow for the use of super resolution techniques.

\bibliography{references} 
\bibliographystyle{spiebib} 

\end{document}